\documentclass[pdflatex,sn-nature,iicol,unnumbered]{sn-jnl}
\usepackage[utf8]{inputenc}
\usepackage{graphicx}
\usepackage[dvipsnames]{xcolor}
\usepackage{epstopdf}
\usepackage{amsmath,amsthm,amssymb}
\usepackage{latexsym}
\usepackage{braket}
\usepackage[normalem]{ulem}
\usepackage[section]{placeins}
\usepackage{mathtools}
\usepackage{multirow}
\usepackage{textcomp,gensymb} 
\usepackage{bm} 
\usepackage{float}
\usepackage{setspace}
\usepackage{array}
\usepackage{changepage}
\usepackage{tikz}
\usepackage{contour}
\usepackage[left]{lineno}
\usepackage{blindtext}
\usepackage{setspace}
\usepackage[none]{hyphenat}
\usepackage{xspace}
\usepackage[mathscr]{euscript}
\newcommand{\hvs}{HoV$_6$Sn$_{6}$}
\newcommand{\dvs}{DyV$_6$Sn$_{6}$}

\newcommand{\rvs}{$R$V$_6$Sn$_{6}$}
\usepackage{hyperref}
\usepackage[tracking=true]{microtype}
\SetTracking{encoding={*}, shape={*}}{20}
\begin{document}
\title{Quartet ferromagnetism in the small-exchange limit}



\author[1,2]{\fnm{Samuel} \sur{Ogundeji}}
\equalcont{These authors contributed equally to this work.}
\author[1,2]{\fnm{Tianxiong} \sur{Han}}
\equalcont{These authors contributed equally to this work.}
\author[1,2]{\fnm{Tyler J.} \sur{Slade}}
\author[5]{\fnm{Olivia} \sur{Medina}}
\author[4]{\fnm{Joanna M.} \sur{Blawat}}
\author[4]{\fnm{Filip} \sur{Ronning}}
\author[4]{\fnm{John} \sur{Singleton}}
\author[4,7]{\fnm{Sean M.} \sur{Thomas}}
\author[6]{\fnm{Alimamy F.} \sur{Bangura}}
\author[3]{\fnm{Wei} \sur{Tian}}
\author[3]{\fnm{Eleanor M.} \sur{Clements}}
\author[3]{\fnm{Bing} \sur{Li}}
\author[3]{\fnm{Adam A.} \sur{Aczel}}
\author[1,2]{\fnm{Paul C.} \sur{Canfield}}
\author[1,2]{\fnm{Benjamin G.} \sur{Ueland}}
\author[1,2]{\fnm{Rebecca} \sur{Flint}}
\author*[1,2]{\fnm{Robert J.} \sur{McQueeney}}\email{rmcqueen@ameslab.gov}
\author*[1,2]{\fnm{Aashish} \sur{Sapkota}}\email{asapkota@ameslab.gov}

\affil[1]{\orgdiv{Division of Materials Sciences and Engineering}, \orgname{Ames National Laboratory, U.S. DOE}, \orgaddress{\city{Ames}, \postcode{50011}, \state{Iowa}, \country{USA}}}

\affil[2]{\orgdiv{Department of Physics and
Astronomy}, \orgname{Iowa State University}, \orgaddress{\city{Ames}, \postcode{50011}, \state{Iowa}, \country{USA}}}

\affil[3]{\orgname{Oak Ridge National Laboratory}, \orgaddress{\street{1 Bethel Valley Road}, \city{Oak Ridge}, \postcode{37830}, \state{Tennessee}, \country{USA}}}

\affil[4]{\orgname{National High Magnetic Field Laboratory, Los Alamos National Laboratory}, \orgaddress{\city{Los Alamos}, \postcode{87545}, \state{New Mexico}, \country{USA}}}

\affil[5]{\orgdiv{Department of Physics}, \orgname{Arizona State University}, \orgaddress{\city{Tempe}, \postcode{85281}, \state{Arizona}, \country{USA}}}

\affil[6]{\orgname{National High Magnetic Field Laboratory}, \orgaddress{\city{Tallahassee}, \postcode{32310}, \state{Florida}, \country{USA}}}

\affil[7]{\orgdiv{Department of Physics},\orgname{Colorado State University}, \orgaddress{\city{Fort Collins}, \postcode{80523}, \state{Colorado}, \country{USA}}}

\date{\today}

\abstract{\mathversion{normal}Rare-earth magnets are often regarded as classical owing to their large angular momentum $J$, but this picture breaks down when exchange is smaller than the crystalline electric-field (CEF) energy scale, allowing the degeneracy of low-lying CEF manifolds to dictate the emergent quantum phases. Here, we uncover such a regime in the $RV_6$Sn$_6$ ($R =$ rare-earth) kagome metals with strong magnetic anisotropy. Thermodynamic and neutron-diffraction measurements reveal successive axial and tilted ferromagnetic (FM) phases in \dvs, arising from independent ordering of the axial $M_z$ and transverse $M_x$ dipolar components enabled by a rare quasi-quartet ground state. On the other hand, \hvs\ has a similar tilted magnetic anisotropy, but remains axial down to $0.03$~K. This discrepancy can be traced back to the non-Kramers nature of Ho$^{3+}$ where the CEF lifts the quartet degeneracy, leaving a quasi-triplet ground state. Unlike quartets or higher multiplets, we find that the effective spin-1 triplet does not allow independent dipolar ordering without strong quadrupolar (nematic) interactions that act to break rotational symmetry, revealing an essential multipolar prerequisite for tilted magnetic ground states. These results establish a quartet CEF manifold as the minimal degeneracy required for dipolar-driven independent ordering. Mapping the surrounding CEF parameter space further reveals competing magnetic, non-magnetic singlet, and potential quadrupolar phases, driven by multifold ground state degeneracies. These results establish CEF tuning as a versatile route to unconventional quantum phases across the broader $RT_6X_6$ ($T$ = transition metal; $X$ = Sn, Ge) family.}

\maketitle

\section{Introduction \label{sec:intro}}
Rare-earth magnets host phenomena ranging from high-performance permanent magnetism\cite{Coey2020} to multipolar order\cite{Morin1990}, frustrated spin liquids\cite{Gardner2010,Rau2019}, and topological magnetic states\cite{Riberolles2022,Riberolles2023,Hung2020Topological}. At the microscopic level, this diverse manifestation is rooted in strongly spin–orbit-coupled, localized $4f$ moments governed by crystalline electric fields (CEFs) and magnetic exchange. Their interplay produces strong single-ion anisotropy\cite{Ke_2015}, whose classical or quantum character depends critically on the relative exchange and CEF energy scales. When exchange dominates, as in rare-earth–transition-metal permanent magnets with strong $3d–4f$ coupling, magnetic anisotropy can be well described classically\cite{Kuzmin1992}. In contrast, when exchange is smaller than the relevant CEF scale, as can occur for weak $4f–4f$ exchange, the discrete symmetry, degeneracy, and wavefunctions of low-lying CEF ground states result in manifestly quantum behaviors.

The degeneracy of this low-energy CEF manifold thus provides a hierarchy of quantum magnetic behavior. An isolated non-Kramers singlet supports only a non-magnetic state \cite{Hasanain1981_PrP}, while a doublet acts as an effective spin-$1/2$, with a single magnetic phase transition and provides the canonical setting for Ising quantum criticality\cite{Bitko1996_LiHoF4,Han_2025_TbV6Sn6}. Triplet manifolds can be described by an effective spin-$1$ degree of freedom that allows a more complex critical behavior captured by Blume-Capel model\cite{Blume1966,Capel1966, QuantumBlumeCapel2015}. Quartets enlarge the local Hilbert space further, opening access to complex magnetic phases with coupled dipolar and multipolar degrees of freedom beyond conventional magnetism\cite{Sundermann_2017_CeB6,Rosenberg2019_YbRu2Ge2_qq,Koga2006_PrOsSb_qq}.

The $RT_6X_6$ family ($R$ = rare earth, $T$ = transition metal and $X$ = Sn, Ge) provides a platform in which both the classical and quantum limits of rare-earth magnetism can be realized within a common structural motifs. In $R$Mn$_6$Sn$_6$, strong coupling between the magnetic rare-earth and Mn sublattices drives the system towards the classical, exchange-dominated limit\cite{Lee_2023,Riberolles2024}, where spin-reorientation (SR) transitions and low-symmetry tilted magnetic structures across the rare-earth series can be largely understood within a classical magnetic-anisotropy framework\cite{Lee_2023}. 

By contrast, in RV$_6$Sn$_6$, the non-magnetic V sublattice leaves substantially weaker $4f$--$4f$ exchange, bringing the discrete structure of the low-lying CEF manifold to the forefront and providing access to the above discussed quantum behavior. Indeed, our studies across the Tb--Dy--Ho series, combining neutron scattering and thermodynamic measurements, establish the quantum nature of their magnetism and demonstrate how the degeneracy of their CEF manifolds dictates their distinct magnetic behaviors. 

Previous work on TbV$_6$Sn$_6$ established an isolated low-energy doublet driving strongly Ising-like ferromagnetism with putative transverse-field Ising quantum criticality\cite{Han_2025_TbV6Sn6}. Here, we show that \hvs\ and \dvs\ instead host low-energy quasi-triplet and quasi-quartet manifolds, respectively. Classically, both compounds are expected to undergo a SR transition from an axial to a tilted ferromagnetic (FM) ground state, as shown in Fig.~\ref{CM}. Experimentally, \dvs\ exhibits an axial-to-tilted FM SR transition with reduced moments relative to classical predictions, whereas \hvs\ remains strictly axial down to the lowest measured temperature. Our mean-field analysis incorporating a quantum treatment of the CEF manifold captures this distinction and identifies the quasi-quartet as the minimal manifold required for the development of independent axial $M_z$ and transverse $M_x$ magnetic order parameters. 

Extending beyond the experimentally determined CEF parameters reveals a rich phase landscape, including competing dipolar and  higher-rank multipolar orders. Particularly intriguing in \hvs\ is a Blume--Capel-like quantum spin-1 regime hosting field-driven quantum phase transitions between axial, planar, and non-magnetic states, near which the experimental parameters reside. More broadly, our extended phase-space mapping places the diverse magnetic ground states observed across the $RT_6X_6$ family within a broader landscape of competing dipolar, multipolar, and quantum-critical regimes, highlighting this chemically tunable family as a fertile platform for exploring CEF-dimensionality-enabled quantum phases.

\section{Results}

$R$T$_6$X$_6$ are hexagonal compounds ($P$6/$mmm$) consisting of FM triangular $R$ layers and $T$ kagome layers\cite{Lee_2022_RV6Sn6}. The $R$ site has a strong CEF potential that drives highly anisotropic magnetism. In DyMn$_6$Sn$_6$ and HoMn$_6$Sn$_6$, easy-cone anisotropy combined with the presence of strong $R$-Mn antiferromagnetic exchange leads to a tilted ferrimagnetic ground state below $\approx 250$~K, as predicted by the classical magnetic anisotropy energy (MAE)\cite{Lee_2023}.  

\begin{figure}[!htbp]
\centering
\begin{adjustwidth}{0mm}{}
\includegraphics[width=0.45\textwidth]{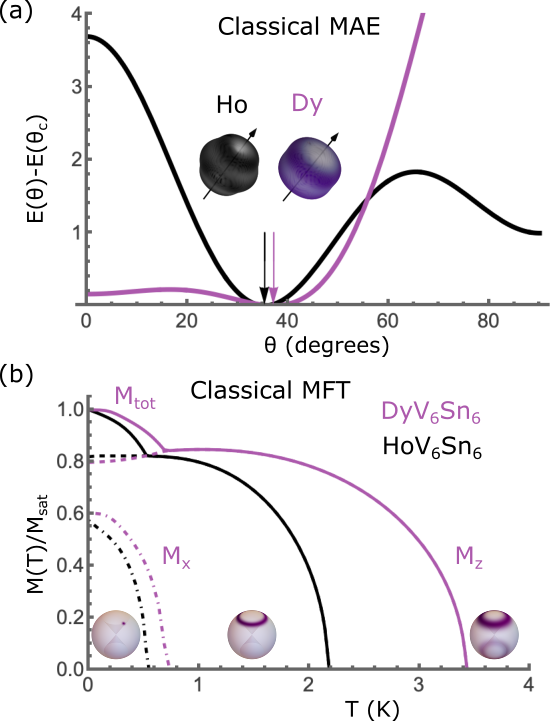}
\caption{\textbf{Classical magnetic anisotropy energy and mean-field results.}(a) The $T = 0$ classical magnetic anisotropy energy, $E(\theta) - E(\theta_c)$ as a function of canting angle, $\theta$ for \hvs\ (black) and \dvs\ (purple), with $\phi =0$ (the minimal direction). The minima are at $\theta_c \approx 35^\circ$ and $\approx 37^\circ$, respectively, as indicated by the arrows. Insets show the $4f$-charge distributions in infinite field along $\theta_c$. (b) Results of mean-field calculations using classical spin-coherent states with magnetic anisotropy constants determined by the experimental CEF parameters and isotropic magnetic exchange chosen to give the highest experimentally observed transition temperature, $T_{\mathrm{C}z}$. The plot shows axial ($M_z$, dashed), transverse ($M_x$, dash-dotted) and total ($M_{tot}$, solid)  magnetization components for \hvs\ (black) and \dvs\ (purple), normalized by the full saturation $(g_L J)$. The insets qualitatively illustrate the solid angle coverage of the spin-coherent states in the three temperature regions, as described in the main text.
\label{CM}}
\end{adjustwidth}
\end{figure}

The same easy-cone anisotropy is also present in DyV$_6$Sn$_6$ and HoV$_6$Sn$_6$ with a classical MAE landscape shown in Fig.~\ref{CM}(a). However, V magnetism is absent and weak $R$-$R$ exchange establishes FM order below $T_\mathrm{C}\approx 3$~K. Classical mean-field theory calculations shown in Fig.~\ref{CM}(b) predict that both compounds first establish axial FM order, followed by a SR transition into a tilted FM ground state (in accordance with the MAE) with fully saturated moment of $\approx 10\mu_\mathrm{B}$. The insets in Fig.~\ref{CM}(b) illustrate the distribution of spin-coherent states on the unit sphere across the three regimes. In the paramagnetic phase, the conical anisotropy produces two degenerate, opposing cones along the $c$~axis that preserve time-reversal $\mathbb{Z}_2$ symmetry with zero net moment. Below the first transition, spontaneous $\mathbb{Z}_2$ symmetry breaking yields an axial FM phase with net $c$-axis polarization while transverse moments fluctuate dynamically around the cone. Classically, further cooling prompts spontaneous ordering of the transverse moment into a tilted phase, with the moments pointing along a single direction on the cone.

Considering the relatively weak exchange, these classical predictions are in direct conflict with reported experimental observations. Magnetization and thermodynamic measurements for \dvs\ and \hvs\ observed the establishment of FM order below $T_\mathrm{C} \approx 3$ and $2.4$~K, respectively~\cite{Lee_2022_RV6Sn6,Zhang_2022_MGS_RV6Sn6_2nd,Zeng_2024_DyHoV6Sn6}. Single-crystal neutron diffraction reports a tilted FM ground state in \dvs, but axial in \hvs~\cite{Zhou_2024__MGS_RV6Sn6_1st}. In both compounds, the ordered moments are substantially reduced relative to their saturated values\cite{Zhou_2024__MGS_RV6Sn6_1st} and no spin reorientation has been reported. However, in \dvs, a specific-heat anomaly near $2.5$~K suggests an additional magnetic phase, consistent with the two transitions predicted by the classical model. In \hvs, a possible low-temperature tilted state remains unexplored, as previous neutron measurements were limited to $T \gtrsim 1.8$~K, while a large nuclear hyperfine contribution obscures the low-temperature heat capacity. 

\begin{figure}[]
\centering
\includegraphics[width=0.49\textwidth]{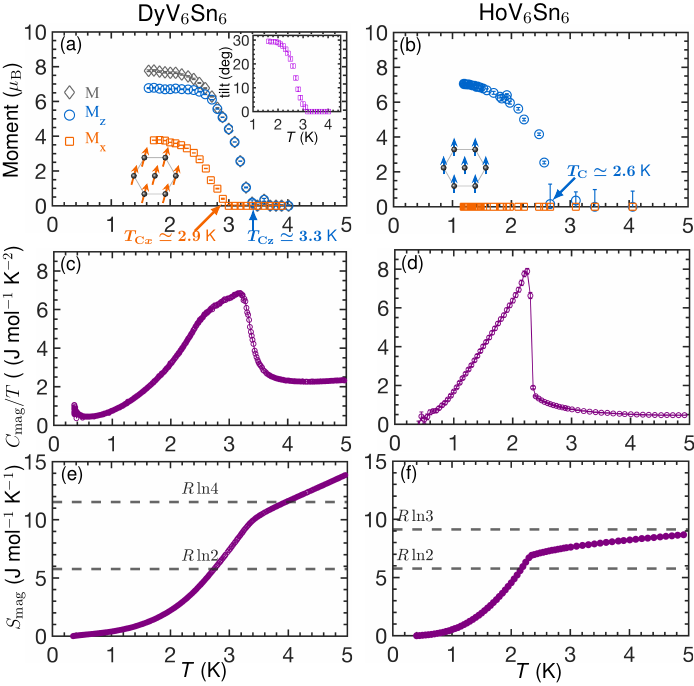}
\caption{\textbf{Temperature evolution of the magnetic ground states in \dvs\ and \hvs.} Temperature dependence of the transverse ($M_x$), axial ($M_z$), and total ($M$) ordered magnetic moments for (a) \dvs\ and (b) \hvs\, extracted from the magnetic-to-nuclear Bragg peak intensity ratios presented in Supplementary Fig.~S3. Their evolution reveals a transition from axial to tilted FM order in \dvs\ at $T_{\mathrm{C}x} \simeq 2.9$~K, whereas \hvs\ retains a c-axis axial FM configuration. The inset in (a) shows the temperature dependence of the tilting angle of the magnetic moments in \dvs\ away from the crystallographic $c$-axis. (c,d) Temperature dependence of the zero-field magnetic heat capacity $C_\mathrm{mag}/T$ for \dvs\ and \hvs, respectively. $C_\mathrm{mag}$ was obtained by subtracting the nonmagnetic lattice contribution of YV$_6$Sn$_6$ reported in Ref.~\cite{Zhang_2022_MGS_RV6Sn6_2nd} over the relevant temperature range($\leq 5$~K). Nuclear Schottky is also subtracted from the data. (e,f) Temperature dependence of magnetic entropy for \dvs\ and \hvs, respectively, obtained from integrating $C_\mathrm{mag}/T$.\label{EEMA}}
\end{figure}

Here, neutron diffraction measurements extending to the lowest accessible temperatures, including the sub-kelvin regime for \hvs, map the complete magnetic phase evolution of both compounds. Figure~\ref{EEMA} (a) and (b) summarizes the neutron diffraction results for \dvs\ and \hvs, respectively. In agreement with classical predictions, \dvs\ exhibits a two-step magnetic ordering: initial axial FM order along the $c$-axis $M_z$ develops at $T_\mathrm{Cz} \simeq 3.3$~K, followed by the onset of a transverse component $M_x$ at $T_{\mathrm{C}x} \simeq 2.9$~K that establishes a tilted FM state. The transverse component manifests as an enhancement of magnetic intensity at $(0,0,L)$ Bragg peaks, such as the $(0,0,2)$ and $(0, 0, 3)$ shown in supplementary Fig.~S2(a)-(b). In contrast, Fig.~S2(c)-(d) shows that the intensities of these peaks for \hvs\ remain unchanged down to $30$~mK, demonstrating that the axial FM state persists to the lowest measured temperatures, with any tilt $\lesssim 6\degree$ remaining undetectable within our experimental uncertainty.

 The heat capacity measurements shown in Figs.~\ref{EEMA}(c) and \ref{EEMA}(d) are consistent with previous reports and our neutron results. For \hvs, the magnetic heat capacity $C_\mathrm{mag}/T$ in Fig.~\ref{EEMA}(d) exhibits a mean-field step consistent with a continuous magnetic phase transition. For \dvs, the primary transition step feature near $3.4$~K develops a pronounced low-temperature shoulder extending to approximately $2.6$~K, suggesting the possibility of a second magnetic transition. Furthermore, the measured magnetic entropy shown in Figs.~\ref{EEMA}(e) and \ref{EEMA}(f) provides insights into the degeneracy of the low-lying CEF states. In \dvs, the entropy approaches $R\ln 2$ near $T_{\mathrm{C}x}$ and $R\ln 4$ above $T_\mathrm{Cz}$, indicating that the successive $M_x$ and $M_z$ order involves four thermally accessible states. In contrast, the entropy of \hvs\ approaches $R\ln 3$ above $T_\mathrm{C}$, indicating three accessible low-energy states.


 Complementary magnetization measurements (Supplementary Information) agree with our neutron results. Together, these measurements map the distinct low-temperature magnetic phase evolutions in \dvs\ and \hvs, demonstrating the failure of classical models and need for a full quantum CEF framework.

To capture the quantum nature of the magnetism, we modeled the temperature dependence of the magnetic ground states using a mean-field Hamiltonian comprising the CEF contribution, $\mathcal{H}_{\mathrm{CEF}}$, and an effective exchange interaction describing FM Heisenberg coupling between rare-earth ions with total angular momentum $J$,
\begin{equation}
\mathcal{H} = \mathcal{H}_{\text{CEF}} - \frac{1}{2} \sum_{ij} \mathcal{J}_{ij} \, \vec{J}_i \cdot \vec{J}_j,
\label{comp_hamil}
\end{equation}

Here, 
$\mathcal{H}_{\mathrm{CEF}} = B_{2}^{0} O_{2}^{0} + B_{4}^{0} O_{4}^{0} + B_{6}^{0} O_{6}^{0} + B_{6}^{6} O_{6}^{6}$, where $\EuScript{O}_m^l$ are the corresponding Stevens operators and $B_l^m$ are the CEF parameters constrained by the hexagonal $D_{6h}$ point-group symmetry of the $R^{3+}$ site. The CEF parameters used in the calculations for \hvs\ were obtained by fitting our unpublished inelastic neutron scattering data. For \dvs, for which no corresponding experimental determination of the CEF parameters is currently available, we employ Stevens scaling of the CEF parameters from those reported for TbV$_6$Sn$_6$ (Case II in Ref.~\cite{Han_2025_TbV6Sn6}) and subsequently refined them to reproduce the experimentally observed magnetic behavior. The CEF parameters used in the calculations are listed in Table~\ref{tab:CEF_parameters_DyHo}. The total effective isotropic exchange coupling, $\mathcal{J}_0 = \sum_j \mathcal{J}_{0j} = z\mathcal{J}_{exch}$ (taking $z = 6$ nearest neighbors), was set to $0.01$ and $0.006$~meV for \dvs\ and \hvs, respectively, to reproduce the experimental Curie temperatures, $T_{\mathrm{C}}$. These values should be regarded as lower-bound estimates of the corresponding interaction strengths. Pertinent energy scales are the exchange field, $h_{\mathrm{exch}} = z\mathcal{J}_{\mathrm{exch}}\langle J_z \rangle$, and the exchange splitting of the low-lying CEF states, $\Delta E_{\mathrm{split}} \sim 2 z\mathcal{J}_{\mathrm{exch}} \langle J_z \rangle^2$.

\begin{table}[]
\centering
\caption{CEF parameters for Dy$^{3+}$ and Ho$^{3+}$, in meV, used in our mean-field analysis, shown in Fig.~\ref{MA_Expdata}.}
\label{tab:CEF_parameters_DyHo}
\begin{tabular}{|c|c|c|c|c|}
\hline
  & $B_{2}^0 \times 10^{2} $ & $B_{4}^0\times 10^{4} $ & $B_{6}^0 \times 10^{6} $ & $B_{6}^6\times 10^{5} $ \\
\hline
Dy$^{3+}$ & $-6.71$ & $2.91$ & $-1.58$ & $-2.49$ \\
\hline
Ho$^{3+}$ & $-1.35$ & $0.62$ & $1.40$ & $-1.82$ \\
\hline
\end{tabular}
\end{table}

\begin{figure}[]
\centering
\includegraphics[width=0.46\textwidth]{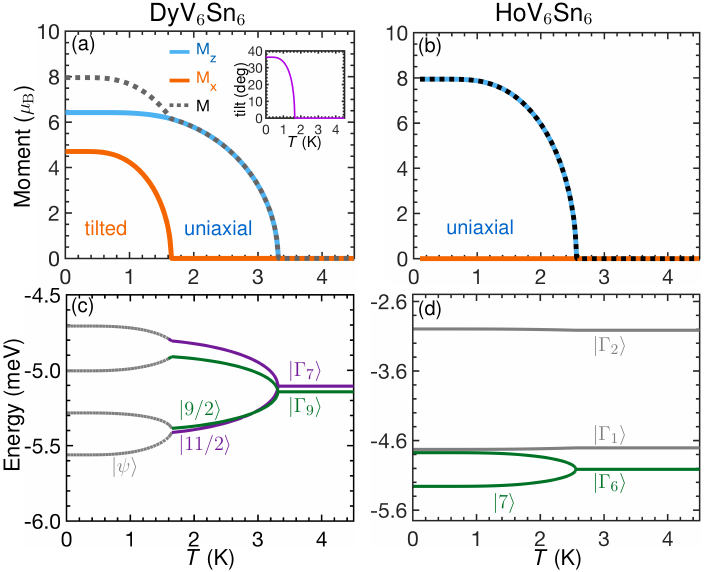}
\caption{\textbf{Quantum mean-field theory results.} (a,b) Magnetic moments obtained from the quantum mean-field calculations for \dvs\ and \hvs, respectively. The insets in (a) show the calculated tilting angles. (c) Calculated evolution of the low-lying CEF levels, highlighting the quasi-quartet in \dvs\ consisting of Kramers doublets $\ket{\Gamma_9}=a\ket{\pm9/2}+b\ket{\mp3/2}+c\ket{\mp15/2}$ and $\ket{\Gamma_7}=f\ket{\pm11/2}+g\ket{\mp1/2}+h\ket{\mp13/2}$ with $a,f\approx1$ and (d) for \hvs\ highlighting the low-lying triplet in \hvs\ above the magnetic transition. The triplet manifold consists of a non-Kramers doublet, $\ket{\Gamma_6}=f\ket{\pm7}+g\ket{\pm1}+h\ket{\mp5}$, and a singlet, $\ket{\Gamma_1}=a\ket{0}+b(\ket{+6}+\ket{-6})$. An additional singlet $\ket{\Gamma_2}=1/\sqrt2(\ket{-6}-\ket{+6})$ exists at higher energy and separated from first singlet by $\approx 1.8$~meV due to the $B_6^6$ CEF pontential.
    \label{MA_Expdata}}
\end{figure}

The resulting CEF-plus-exchange Hamiltonian was then solved self-consistently to determine the temperature evolution of the magnetization, with the results summarized in Fig.~\ref{MA_Expdata}. The mean-field model qualitatively captures the distinct magnetic behavior of the two compounds, reproducing the successive axial-to-tilted FM ordering in \dvs\ [Fig.~\ref{MA_Expdata}(a)] and the persistence of the axial FM state in \hvs\ [Fig.~\ref{MA_Expdata}(b)], as well as the experimentally observed magnitudes of the ordered moments. Some discrepancies nevertheless remain, most notably in the calculated transition temperature of the tilted phase and, to a lesser extent, in the maximum value of the tilting angle. These deviations likely reflect interactions beyond the minimal CEF-plus-isotropic-exchange model, such as long-range dipolar interactions or anisotropic exchange terms. An extended treatment incorporating anisotropic exchange interactions is presented in the Supplementary Information (Section~S2; Fig.~ S4) and yields results in closer quantitative agreement of transition temperatures with the experimental observations. Nevertheless, the minimal model discussed here captures the essential physics and allows for a conceptual understanding based on the ground state CEF manifold.

For \dvs\ ($J = 15/2)$, as shown in [Fig.~\ref{MA_Expdata}(c)], the paramagnetic regime above $T_\mathrm{Cz}$ consists of two nearly degenerate Kramers doublets, $\ket{\Gamma_9} \approx \ket{\pm 9/2}$ and $\ket{\Gamma_7} \approx \ket{\pm 11/2}$. Furthermore, the CEF doublet splitting is 0.04 meV which is much smaller than the estimated exchange splitting of $\approx 0.6$~meV, forming an effective quasi-quartet. Below $T_\mathrm{Cz}$, the self-consistent exchange field breaks time-reversal symmetry and lifts the Kramer's degeneracy of both doublets, stabilizing the axial FM phase that retains a low-energy quasi-doublet composed of $\ket{+9/2}$ and $\ket{+11/2}$. These ``axial'' magnetic phases for eigenstates $m_\mathrm{J} < J$ represent the quantum analog of the above discussed classical transversely fluctuating easy-cone magnetic state. The quasi-doublet possesses a large off-diagonal matrix element $\bra{+11/2} J_x \ket{+9/2}$ and a small energy splitting. Consequently, the transverse Van Vleck susceptibility diverges upon cooling, driving order of the transverse moment $M_x$. The self-consistent CEF scheme in Fig.~\ref{MA_Expdata}(c) shows that, in the tilted phase, the ground-state wavefunction is described by $\ket{\psi} \approx \alpha\ket{+9/2} + \beta\ket{+11/2}$, yielding simultaneous non-zero $M_z$ and $M_x$ moments with a reduced total magnetic moment of $\sim 8 \mu_\mathrm{B}$.

In contrast, the non-Kramers nature of \hvs\ allows the CEF to lift the four-fold degeneracy leaving a low-energy quasi-triplet, comprising a non-Kramers $\ket{\Gamma_6} \approx \ket{\pm 7}$ doublet and a nearby $\ket{\Gamma_1}=a\ket{0}+b(\ket{+6}+\ket{-6})$ singlet. The low-lying triplet is consistent with the magnetic entropy results shown in Fig.~\ref{EEMA}(f). Upon cooling through the magnetic transition, the self-consistent exchange field lifts the degeneracy of the $\Gamma_6$ doublet and $M_z$ orders. The $\Gamma_1$ singlet energy remains nearly constant in the axial exchange field, indicating weak mixing with the $\Gamma_2$ singlet and other CEF levels. The isolation of the quasi-triplet justifies its treatment as an effective spin-1 system. 

Although $J_x$ mixes the singlet with the exchange-split doublet, the van Vleck susceptibility always remains subcritical due to the increasing gap size and the transverse ordered moment cannot develop. This is a general property of spin-1 Blume-Capel model where axial, planar, or non-magnetic phases cannot coexist. Generically, dipolar interactions can stabilize either purely axial or planar magnetic order, possibly with a first-order transition between the two\cite{Fridman_2008}. However, adding moderate strength quadrupolar interactions can stabilize the tilted phase, where in-plane dipolar and shear quadrupolar interactions can work together to order both $\langle O_{xz}\rangle$ and $\langle J_x\rangle$ as a second order transition not possible with purely dipolar or purely quadrupolar interactions.

\section{Summary and Discussion}

Our experimental results establish the complete low-temperature magnetic phase evolution of \dvs\ and \hvs, uncovering distinct evolution of their magnetic anisotropy; axial to tilted FM in \dvs\ and strictly axial down to sub-kelvin temperature in \hvs. Our analyses highlight the importance of the degeneracy of the low-lying CEF manifolds. We find that a quasi-quartet is the minimal degeneracy needed for independent ordering of axial and transverse magnetization components -- enabling a continuous spin-reorientation transition into a tilted FM phase -- a phenomenon that we term quartet ferromagnetism. Crucially, the quasi-quartet in \dvs\ consists of two Kramers doublets whose wavefunctions differ by $\Delta m = 1$, allowing both the axial (Curie) and transverse (van Vleck) susceptibilities to diverge, leading to independent $M_z$ and $M_x$ order. By contrast, in \hvs, the quasi-triplet ground state suppresses the transverse order and stabilizes axial ferromagnetism down to sub-kelvin temperatures.

\begin{figure*}[]
\centering
\includegraphics[width=0.95\textwidth]{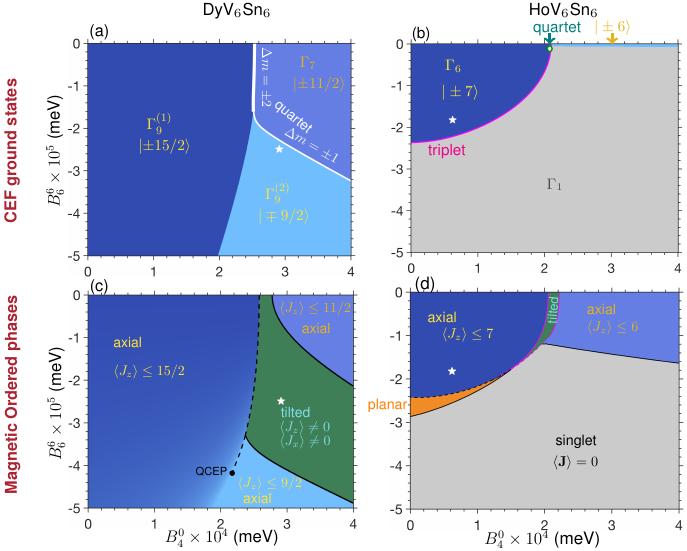}
\caption{\textbf{Phase maps of the low-energy CEF and magnetic states.} Two-dimensional phase diagrams as a function of the CEF parameters $B_{66}$ and $B_{40}$ for \dvs\ [(a,c)] and \hvs\ [(b,d)]. The white stars mark the optimized/experimental CEF parameters listed in Table~\ref{MA_Expdata} and used in the calculations discussed above. (a,b) Single-ion phase maps obtained from $H_{\mathrm{CEF}}$ illustrating the CEF ground states. \dvs\ in (a) comprises three regions distinguished by their dominant $\ket{m_J}$ character but belonging to only two irreducible representations, $\Gamma_9$ and $\Gamma_7$. $\Gamma_9$ is further defined as $\Gamma_9^{(1)}$ and $\Gamma_9^{(2)}$ to distinguish the dominant $\ket{15/2}$ and $\ket{9/2}$ states. The white boundaries mark quartet degeneracies formed by the crossing of the $\ket{11/2}$- and $\ket{9/2}$-dominated doublets ($\Delta m=1$), while thicker white line is the quartet with $\ket{15/2}$- and $\ket{11/2}$-dominated doublets ($\Delta m=2$). In \hvs\ (b), the quartet degeneracy is restricted to a single point at $B_{66} \approx 0$. For B66 $\neq 0$,the phase boundary (magenta line) between the doublet $\Gamma_6$ and singlet $\Gamma_1$ has a triplet degeneracy. (c,d) Magnetic ordered phases obtained from the self-consistent solution of $H=H_{\mathrm{CEF}}+H_{\mathrm{ex}}$ at $T = 0$~K. The phases are classified by the axial and transverse moments as axial ($J_x=0$; shades of blue), tilted($J_z\neq0$, $J_x\neq0$;; green), easy-plane ($J_z=0$, $J_x\neq0$; orange), or nonmagnetic/singlet ($\mathbf {J}=0$; gray). Dotted dark/magenta colored phase boundaries denote first-order transitions, whereas solid boundaries denote second-order transitions. For \dvs, the first-order transition line in (c) terminates at a quantum critical end point (QCEP), beyond which the two symmetry-equivalent axial FM phases evolve continuously into one another. In (d), for \hvs, the tilted and planar phases taper into narrow, wedge-like regions separating the axial and singlet phases. Near these tapered ends, a small variation in $B_4^0$ drives a succession of four distinct ground states: axial, tilted, planar and singlet.
\label{PM}}
\end{figure*}

The central role of the quasi-quartet in stabilizing the tilted phase is illustrated by the CEF phase maps in Fig.~\ref{PM}. In the phase maps, we vary two key CEF parameters; B$_{4}^0$, which primarily controls the tilt angle, and B$_{6}^6$ which lifts the degeneracy of some non-Kramers doublets.  Other model parameters are held fixed. The phase diagrams obtained from the single-ion CEF Hamiltonian ($H_{\rm CEF}$) and mean-field models ($H_{\rm CEF}+H_{\rm exchange}$) are shown in Figs.~\ref{PM}(a,b) and (c,d), respectively. For \dvs, the single-ion phase map reveals three prominent regions consisting of the $\Gamma_9^{(1)} \approx \ket{\pm 15/2}$, $\Gamma_9^{(2)}\approx \ket{\pm 9/2}$ and $\Gamma_7 \approx \ket{\pm 11/2}$ doublets. At the intersection of the $\Gamma_9$ and $\Gamma_7$ regions, a level crossing generates a line of quartet degeneracy.  Such low-lying quartet or quasi-quartet manifolds are of interest because they can drive exotic multipolar phenomena, including antiferroquadrupolar and hidden order in CeB$_6$\cite{Sundermann_2017_CeB6,Koitzsch2016_hiddenorder}, quadrupolar correlations near unconventional superconductivity in PrOs$_4$Sb$_{12}$\cite{Koga2006_PrOsSb_qq,Sato2010_PrOsSb_qq}, ferroquadrupolar order in YbRu$_2$Ge$_2$\cite{Rosenberg2019_YbRu2Ge2_qq}, and coupled dipolar--quadrupolar order in DyAuGe\cite{Kurumaji2025_DyAuGe_qq}. However, for the \dvs\ CEF parameters [the star in Fig.~\ref{PM}(a)], the dipolar ordering channels are dominant. This ordering is controlled by the $\Gamma_9^{(2)}$ and $\Gamma_7$ quartet, whose symmetry permits interdoublet mixing via $J_x$ and $J_y$ ($\Delta m = \pm 1$), and therefore provides the microscopic prerequisite for developing a transverse dipolar moment. Interestingly, $\Gamma_9^{(1)}$-$\Gamma_7$ quartet symmetry allows interdoublet mixing via quadrupolar interactions ($\Delta m = \pm 2$). We point out that the $\Gamma_9^{(1)}-\Gamma_9^{(2)}$ boundary is an avoided crossing between two CEF states with the same symmetry and thus does not form a quartet.

The extended $\Delta m = \pm1$ quartet line delineates a broad regime where exchange-driven mixing of the two CEF doublets can yield a ground-state wavefunction supporting concurrent axial ($M_z$) and transverse ($M_x$) moments. Indeed, upon introducing exchange, [Fig.~\ref{PM}(c)], this boundary expands into a finite region of tilted ferromagnetism, within which the optimized CEF parameters for \dvs\ (Table~\ref{tab:CEF_parameters_DyHo}) lies. Although dipolar magnetism dominates in \dvs\ without direct evidence for multipolar order, the finite calculated quadrupolar expectation values, such as $\langle O_2^0\rangle=\langle 3J_z^2-J(J+1)\rangle$ and $\langle O_{xz}\rangle=\langle J_xJ_z+J_zJ_x\rangle$, demonstrate that the quasi-quartet in \dvs\ inherently harbors multipolar degrees of freedom.

The phase map of \hvs\ reveals two regions consisting of $\Gamma_6 \approx \ket{\pm 7}$ doublet and the $\Gamma_1$ singlet.  The singlet arises from the splitting of the $\ket{\pm 6}$ non-Kramers doublet by the $B_6^6$ CEF parameter. Thus, a quartet state occurs only at the isolated $B_{6}^6=0$ point, whereas for $B_{6}^6\neq0$ the boundary corresponds to a triplet manifold formed by the degenerate $\Gamma_6$ doublet and $\Gamma_1$ singlet. The extended quartet manifold associated with the tilted phase in \dvs\ is absent in \hvs, providing the key distinction. Upon including exchange [Fig.~\ref{PM}(d)], a tilted FM phase region does appear only near $B_{6}^6=0$ and is associated with quartet ferromagnetism. However, the experimental parameters lie well within the axial region. The triplet line hosts neighboring axial, planar, and non-magnetic singlet regimes that are consistent with an effective spin-1 Blume--Capel model\cite{Capel1966, Jensen1991} connected by field driven quantum-critical behavior. Further calculations for an effective spin-1 model, presented in the Supplementary Information support this analogy.


The proximity to dipolar and multipolar instabilities, and novel quantum critical behavior, highlights \rvs\ systems as promising candidates for experimental tuning. While individual CEF parameters cannot be easily or predictably varied, applied strain\cite{Jensen1991} or targeted chemical substitution offers viable pathways to modify the local crystal electric field, providing a strategy to access these nearby magnetic and multipolar phases. More broadly, the diverse magnetic signatures reported across $R T_6 X_6$ compounds ($T = 3d, 4d$ transition metals; $X = \text{Sn}, \text{Ge}$)\cite{Ortiz2025,Yang_2024,Ishii_YCr6Ge6} provide evidence that chemical substitution is indeed an effective knob for tuning these quantum states.

\section*{Methods}
\subsection{Sample Growth and Characterization}
Using the fractionation method \cite{Canfield2025FritDisc,Slade2022Eutectic}, we found that both \dvs\ and \hvs\ crystallizes between approximately $775-1025 ^\circ$C and that the following two-step growth sequence is optimal for producing large single crystals. Elemental pieces of Dy/Ho (Ames Lab, $99.99\%$), V (Alpha-Aesar, $99.99 \%$) and Sn (Alpha-Aesar, $99.99\%$) were weighed in a $1:6:60$ molar ratio and loaded into the growth side of a 5 ml alumina Canfield crucible set\cite{Canfield2016FritDisc}. The crucibles were sealed under vacuum in a fused silica ampule and heated in a box furnace to $1180 ^\circ$C. After holding at $1180 ^\circ$C for $\approx 8$~h, the samples were cooled to $1025 ^\circ$C, upon which the liquid phase is cleanly captured in the second crucible on the top side of the crucible set. This first step allows us to both maximize the quantity of V in the growth solution and cleanly separate the liquid from any undissolved V, which can serve as nucleation sites and limit the size of the crystals. A second sequence is then carried out, with the captured decant used in the “growth” side of a new crucible set. The second crucible set was sealed under vacuum as before and warmed in a box furnace to $1050 ^\circ$C. After holding at $1050 ^\circ$ for $6$~h, the furnace was cooled to $775 ^\circ$C over 200 h, upon which the ampule was removed from the furnace and the excess liquid decanted. Once cool, the ampules and crucibles were opened to reveal relatively large (10-20 mg), single phase, crystals of \dvs\ or \hvs. The stoichiometric ratio of 166 was confirmed by energy dispersive X-ray spectroscopy (EDS) and single-crystal X-ray diffraction measurements.

\subsection{Magnetization}
DC magnetization measurements were carried out using a Quantum Design Magnetic Property Measurement System (MPMS Classic) superconducting quantum interference device (SQUID) magnetometer, operated over the temperature range $1.8~\mathrm{K} \leq T \leq 300~\mathrm{K}$ under an applied magnetic field of $H=1~\mathrm{kOe}$ for \hvs\ and $1.8~\mathrm{K} \leq T \leq 30~\mathrm{K}$ under an applied magnetic field of $H=50~\mathrm{Oe}$ for \dvs. Both zero-field-cooled (ZFC) and field-cooled (FC) protocols were used. Each sample was measured with the magnetic field applied both parallel and perpendicular to the hexagonal $c$ axis. The samples were glued onto a Kel-F disk and placed inside a plastic straw. The contribution of the disk to the measured magnetic moment was independently measured and subtracted from the data.

High-field magnetization measurements were carried out at the Pulsed-Field Facility of the National High Magnetic Field Laboratory in Los Alamos. Magnetization was recorded with a compensated-coil extraction magnetometer\cite{Goddard_2008}. An oriented crystal fragment is mounted in an ampoule that can be shuttled in and out of the. Because the pulsed field H(t) varies in time, the voltage induced in the coil is proportional to dM/dt. M(H) is obtained by integrating the signal taken with the sample in the coil and subtracting the integrated signal from an empty-coil run.

\subsection{Thermodynamic measurements}
The ac-heat capacity of \dvs\ was measured in the superconducting magnets at the NHMFL in Tallahassee, using high-resolution differential SiNx membrane-based nanocalorimeters \cite{Tagliati2012}. Each differential nanocalorimeter has two nearly identical devices, one for the sample and one for the reference. Each device is a stack of electrically isolated layers separated by insulating oxide: a thermometer, an AC-driven heater and an offset heater. The sample sits directly on the stack and is thermally anchored to it with Apiezon N grease. A DC current through the offset heater sets and controls the sample temperature\cite{Khansili2023}. The sample was prepared with a Ga-ion focused ion beam (FIB), which cut a lamella with dimensions of approximately $50 \times 20 \times 3 \mu m^3$. The heat capacity of \hvs\ was measured from 0.4 to 5 K using a Quantum Design Physical Property Measurement System (PPMS), which employs a quasi-adiabatic thermal relaxation method. 

\subsection{Single-crystal X-ray diffraction}
Single-crystal x-ray diffraction was performed using a Rigaku XtaLab Synergy-S diffractometer with Ag radiation ($\lambda = 0.56087\,\mathrm{\AA}$), in transmission mode, operating at 65~kV and 0.67~mA. The samples were held in a nylon loop with Apiezon N grease. The total number of runs and images
was based on the strategy calculation from the program \textit{CrysAlisPro} (Rigaku OD, 2023). The data integration and reduction were also performed using \textit{CrysAlisPro}, and a numerical absorption correction was applied based on Gaussian integration over a face-indexed crystal. The structures were
solved by intrinsic phasing using the SHELXT software package and were refined with SHELXL.

\subsection{Neutron Diffraction}
The neutron diffraction measurements were performed at the TAX and VERITAS triple-axis spectrometers at the High Flux Isotope Reactor, Oak Ridge National Laboratory. \dvs\ was measured at TAX and \hvs\ at VERITAS. For the measurements, a plate-like crystal of mass $14.3$~mg of \dvs\ and $18$~mg of \hvs\ were mounted on an aluminum and copper plate, respectively, and aligned in the $(H, 0, L)$ horizontal scattering plane. 

At TAX, the Al holder with the \dvs\ single crystal was attached to a cold finger of the orange cryostat with the temperature range of $1.5-315$~K. The measurements were carried out with fixed final energy of $E_\mathrm{f} = 35$~meV and horizontal beam collimations $48^\prime$--$40^\prime$--$40^\prime$--$120^\prime$. Two PG filters, after the sample, were used to remove neutrons from the higher-order reflections.

At VERITAS, the Cu holder with the \hvs\ single crystal was attached to the dilution fridge insert and the measurements down to $0.03$~K was carried out. The measurements were carried out with fixed initial energy of $E_\mathrm{i} = 14.45$~meV and horizontal beam collimations $40^\prime$--$40^\prime$--$40^\prime$--$80^\prime$. Two PG filters, before the sample, were used to remove neutrons from the higher-order reflections.

\subsection{Classical mean-field theory}

We treat our classical moments as unit vectors defined by their $\theta$ and $\phi$ and write the effective Hamiltonian,
\begin{align}
    H(\theta,\phi,M_x, M_z) & = K_1 \sin^2\theta + K_2 \sin^4 \theta  \cr 
    & + (K_3 + K_3' \cos 6\phi) \sin^6 \theta \cr
    &- \lambda_x M_x \sin \theta \cos \phi -\lambda_z M_z\cos \theta
\end{align}
$(K_1, K_2, K_3,K_3') = (1.7, -12.8, 20.6,-1.4)$meV for DyV$_6$Sn$_6$ and $(K_1, K_2, K_3,K_3') = (-25.5, -53.6, -29.1,-1.6)$meV for HoV$_6$Sn$_6$.  
As the minima are along the $\hat x$ axis here, we set $
M_y = 0$ without loss of generality.
This classical mean-field approach accounts for the effective temperature dependence of anisotropy terms, as treated by Callen-Callen theory \cite{Callen1965}, where $\lambda_x, \lambda_z$ are $\propto J_{XX,ZZ}$.  We can then solve for $M_x, M_z$ self-consistently, 
\begin{align}
    M_x & = 1/Z\int d\Omega \mathrm{e}^{-\beta H(\theta,\phi, M_x,M_z)} \sin \theta \cos \phi \cr 
    M_z &= 1/Z\int d\Omega\mathrm{e}^{-\beta H(\theta,\phi,M_x,M_z)} \cos \theta
\end{align}
where $Z = \int d\Omega \mathrm{e}^{-\beta H}$ and $\beta = 1/(k_B T)$.  We adjust $\lambda_z = 4.8$K or $3.5$K to match the experimental $T_z$ for DyV$_6$Sn$_6$ and HoV$_6$Sn$_6$, respectively. The experimental $T_x$ can be obtained by increasing $\lambda_x$, as
\begin{equation}
    T_x/T_z = \lambda_x/\lambda_z \tan^2 \frac{\theta_c}{2},
\end{equation}
where $\theta_{c,Dy} = 37.1^\circ$ and $\theta_{c,Ho} = 35.3^\circ$.  We can multiply $M$ by $g_L J$ to get the magnetization, which saturates at $\approx 10\mu_B$ for both materials, as shown in Fig.~\ref{CM}.  Reproducing the experimental $T_x/T_z$ for DyV$_6$Sn$_6$ requires strongly planar interactions, $\lambda_x = 3.5 \lambda_z$, while even isotropic interactions for HoV$_6$Sn$_6$ yield $T_\perp \approx .5$K, which is inconsistent with the experiment.


\section*{Acknowledgements}
Work at Ames National Laboratory was supported by the U.\,S.\ Department of Energy (DOE), Basic Energy Sciences, Division of Materials Sciences \& Engineering, under Contract No.\ DE-AC$02$-$07$CH$11358$. A portion of this research used resources at the High Flux Isotope Reactor, a DOE Office of Science User Facility operated by the Oak Ridge National Laboratory. The beam time was allocated to TAX and VERITAS instruments on proposal numbers IPTS-35396, IPTS-36819, and IPTS-34247. A portion of this work was performed at the National High Magnetic Field Laboratory, which is supported by National Science Foundation Cooperative Agreement No. DMR-2128556 and the State of Florida. This work was performed, in part, at the Center for Integrated Nanotechnologies, an Office of Science User Facility operated for the U.S. Department of Energy (DOE) Office of Science. Los Alamos National Laboratory, an affirmative action equal opportunity employer, is managed by Triad National Security, LLC for the U.S. Department of Energy’s NNSA, under contract 89233218CNA000001. Heat capacity measurements of HoV6Sn6 were supported by the U.S. DOE, Basic Energy Sciences, Division of Materials Sciences \& Engineering ``Quantum fluctuations of narrow band systems" project. 

\section*{Author contributions}
A.S. and R.J.M. conceived the study. S.O., T.H., A.S., B.G.U., and R.J.M. planned and conducted the neutron-scattering experiments at ORNL, with support from the experimental teams at the VERITAS (A.A.A., W.T., and B.L.) and TAX (E.M.C.) beamlines. Visualization and analysis of the neutron-scattering data were carried out by S.O., A.S., and T.H., in discussion with R.J.M. and B.G.U. Heat capacity measurements were performed at Los Alamos National Laboratory by  S.M.T. and F.R., and at National High Magnetic Field Laboratory facilities in Tallahassee, Florida by A.F.B. and J.M.B. Pulsed-field magnetization measurements were carried out by J.S. and J.M.B. Bulk single-crystal growth, sample characterization, and magnetization measurements were carried out by T.J.S. and P.C.C. Theoretical modeling was performed by A.S., O.M., R.J.M., and R.F. O.M. performed pre-experimental theoretical modeling to refine the CEF parameters for \dvs\ used in our analysis. A.S. and T.H. analyzed and modeled the inelastic neutron scattering data of \hvs\ to determine the CEF parameters used in the manuscript. Authors A.S., R.J.M., and R.F. drafted the manuscript with input from all authors.

\section*{Correspondence and requests for materials}
Correspondence and requests for materials should be addressed to the corresponding authors.

\onecolumn

\begin{center}
    \vspace*{1cm}
    {\Huge\bfseries Supplementary Information}
\end{center}

\vspace{1cm}

\begin{appendices}

\section{Experimental Results: \label{ER}}
This section contains additional experimental results from magnetization, single crystal x-ray and neutron diffraction.
\subsection{Single crystal x-ray diffraction:}

Although the growth of \textit{R}166 single crystals is well established, single-crystal x-ray diffraction measurements were performed to verify the stoichiometry of the synthesized crystals. This was done to rule out any vacancies or disorder, that can consequentially have effects on the CEF parameters and magnetism. Structural determination or refinement yielded the room-temperature lattice parameters, atomic positions, site occupancies, and space group. The refined crystal structure adopts the $P6/mmm$ space group, consistent with previous reports. Within the experimental uncertainty, no appreciable site vacancies were detected. The refined structural parameters are summarized in the tables below.

\begin{table*}[]
\caption{Single crystal data and structural refinement information at room temperature.}
\begin{tabular}{ccc}
\hline
 Refined formula  & DyV$_{6}$Sn$_{6}$  &  HoV$_{6}$Sn$_{6}$\\
 Formula weight (g/mol) &   $1180.28$ &1182.71  \\
 Crystal system  & Hexagonal  & Hexagonal\\
 Space group, Z & P$6$/mmm, 1 & P$6$/mmm, 1 \\
 a = b (\AA) & $5.5196 (3)$ & 5.5186(2) \\
 c (\AA) & $9.1828 (5)$   & 9.1854(5)\\
 $\alpha = \beta (^\circ)$ & 90  & 90\\
 $\gamma (^\circ)$ & 120  & 120\\
 Crystal size (mm$^3$) & $0.066 \times0.047\times0.044$ & $0.08\times0.065\times0.025$  \\
 Volume (\AA$^3$) & $242.28 (3)$  & $242.26(2)$ \\
$ \rho$ (gcm$^{-3}$) & $8.089$ & $8.107$\\
 $ 2\theta$ range for data collection ($^\circ$) & $3.5 - 63.648$ & 5.98$^{\circ}$ to 62.958$^{\circ}$ \\
\multirow{3}{*}{Index ranges}
& $-9 \leq h \leq 9$ & $-6 \leq h \leq 7$\\
& $-7 \leq k \leq 10$ & $-7 \leq k \leq 6$ \\
& $-15 \leq l \leq 17$ & $-16 \leq l \leq 19$ \\
 Reflections collected & $6497$ & 1951 \\
 Independent reflections & $378$  & 187\\
 $R_{int}$ & $0.0239$ & 0.0377 \\
 Parameters & 20 & 8\\
 $R_1, wR_2 [I \geq 2 \sigma(I)]$ & $ 0.0089$, $ 0.0204$ & $ 0.016$, $ 0.0421$  \\
 Goodness of fit & $1.118$ & 1.228 \\
 Diffraction peak and hole (e \AA$^{-3}$) & $1.50$, $-1.46$ & $0.73$, $-9.96$ \\
 \hline
\end{tabular} 
\end{table*}

\begin{table}[t]
\centering
\caption{Atomic coordinates for DyV$_6$Sn$_6$ at $293.15$\,K.}
\label{tab:DyV6Sn6_coords}
\begin{tabular}{ccccc}
\hline\hline
Atom & Wyckoff & $x$ & $y$ & $z$ \\
\hline
Dy01 & $1a$ & 0.00000 & 0.00000 & 0.00000 \\
Sn02 & $2e$ & 0.00000 & 0.00000 & 0.33319(2) \\
Sn03 & $2c$ & 0.33333 & 0.66667 & 0.00000 \\
Sn04 & $2d$ & 0.33333 & 0.66667 & 0.50000 \\
V05  & $6i$ & 0.50000 & 0.00000 & 0.24835(2) \\
\hline\hline
\end{tabular}
\end{table}

\begin{table}[t]
\centering
\caption{Atomic coordinates for HoV$_6$Sn$_6$ at $293.15$\,K.}
\label{tab:HoV6Sn6_coords}
\begin{tabular}{ccccc}
\hline\hline
Atom & Wyckoff & $x$ & $y$ & $z$ \\
\hline
Ho01 & $1a$ & 0.00000 & 0.00000 & 0.00000 \\
Sn02 & $2e$ & 0.00000 & 0.00000 & 0.33327(3) \\
Sn03 & $2c$ & 0.33333 & 0.66667 & 0.00000 \\
Sn04 & $2d$ & 0.33333 & 0.66667 & 0.50000 \\
V05  & $6i$ & 0.50000 & 0.00000 & 0.24822(4) \\
\hline\hline
\end{tabular}
\end{table}

\begin{table}[]
\centering
\caption{ Anisotropic and equivalent isotropic displacement parameters for DyV$_{6}$Sn$_{6}$ at $293.15$ \, K}
\begin{tabular} {ccccc}
 \hline
   Atoms  &  $U_{11}$ ($10^{-3}$ \AA$^2$)  & $U_{22}$ ($10^{-3}$ \AA$^2$) &  $U_{33}$ ($10^{-3}$ \AA$^2$) \\
 \hline
 Dy$01$  &  $5.18(6)$  & $5.18(6)$ & $8.39(8)$\\
 Sn$02$ &  $5.27(6)$  & $5.27(6)$ &$8.44(8)$\\
 Sn$03$ &  $5.39(6)$  & $5.39(6)$ & $3.84(8)$\\
 Sn$04$ &   $6.92(6)$  & $6.92(6)$ & $3.65(8)$\\
 V$05$ &  $5.13(9)$  & $5.26(11)$ & $3.91(11)$\\
 \hline
\end{tabular} 
\end{table}

\subsection{Magnetization results}

\begin{figure*}[]
\centering
\includegraphics[width=0.95\textwidth]{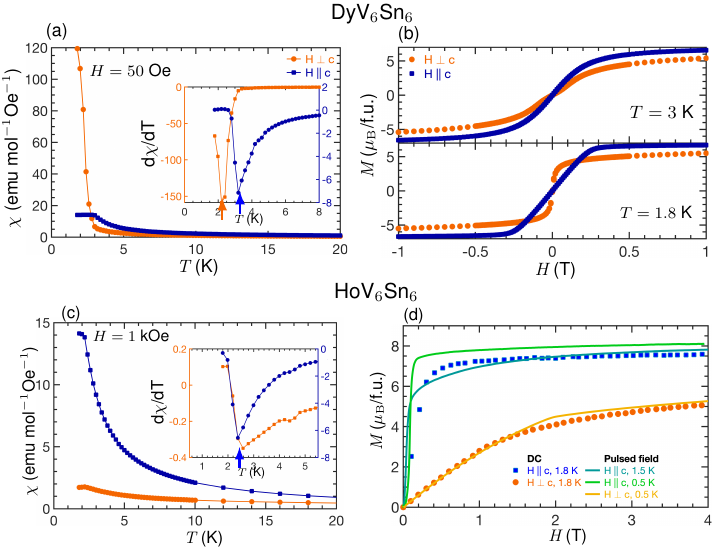}
\caption{(a) and (c) Temperature dependence of the zero-field-cooled magnetic susceptibility of \dvs\ and \hvs, respectively, measured with magnetic fields applied parallel ($H \parallel c$) and perpendicular ($H \perp c$) to the crystallographic $c$-axis. The measurements were performed under applied fields of 50~Oe for \dvs\ and 1~kOe for \hvs. The insets show the temperature derivatives of the susceptibility, highlighting two magnetic phase transitions in \dvs\ and a single transition in \hvs, as indicated by arrows. (b) Field-dependent magnetization of \dvs\ measured at 1.8 and 3~K, revealing a pronounced change in magnetic anisotropy at lower temperature. (d) Field-dependent magnetization of \hvs\ measured using DC fields at 1.8~K and pulsed magnetic fields (solid lines) down to 0.5~K. The results shows no change in magnetic anisotropy down to 0.5 K.\label{MvsHT}}
\end{figure*}

The occurrence of two consecutive magnetic transitions--first into a $c$-axis axial ferromagnetic (FM) state followed by a tilted FM phase in \dvs, compared to a single transition to axial FM phase in \hvs--is further substantiated by our magnetization data (Fig.~\ref{MvsHT}). The temperature derivative of the magnetic susceptibility, shown in inset, for both directions displays minima at distinct temperatures for \dvs, whereas \hvs\ exhibits only a single minimum. Furthermore, isothermal magnetization curves measured at various temperatures [Fig.~\ref{MvsHT}(b, c)] clearly demonstrate an anisotropy change from the $c$-axis toward an easy-plane in \dvs\ upon cooling, while \hvs\ remains strictly uniaxial down to the base temperature of $0.5$~K. These findings are fully consistent with the experimental observations and analysis presented in the main text.

\subsection{Neutron diffraction Results}
\begin{figure*}[]
\centering
\begin{center}
	
\end{center}\includegraphics[width=0.95\textwidth]{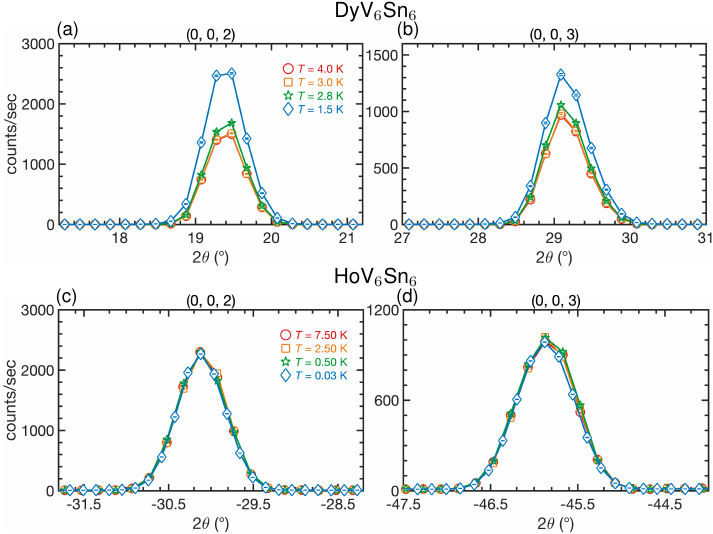}
\caption{(a)-(d) Temperature evolution of the $(0,0,L = 2, 3)$ Bragg peaks for \dvs, measured with fixed final energy of 35 meV (a)-(b) and \hvs, measured with incident neutron energy of 14.45 meV (c)-(d).\label{ND}}
\end{figure*}

Figure~\ref{ND} shows the temperature evolution of the (0, 0, 2) and (0, 0, 3) Bragg peaks for \dvs\ [panels (a)–(b)] and \hvs\ [panels (c)–(d)]. In the tilted FM phase, the development of an in-plane magnetic moment component is expected to enhance the intensities of the (0, 0, L) Bragg reflections through an additional magnetic scattering contribution. Consistent with the emergence of a tilted FM phase, both reflections in \dvs\ exhibit an additional intensity enhancement below $T_{\mathrm{C}x} \sim 2.9$~K. In contrast, no comparable enhancement is observed in \hvs\ down to 0.03~K, suggesting the absence of any such tilting, and confirming the c-axis axial FM ordering as discussed in the main text.

\begin{figure*}[]
\centering
\includegraphics[width=0.95\textwidth]{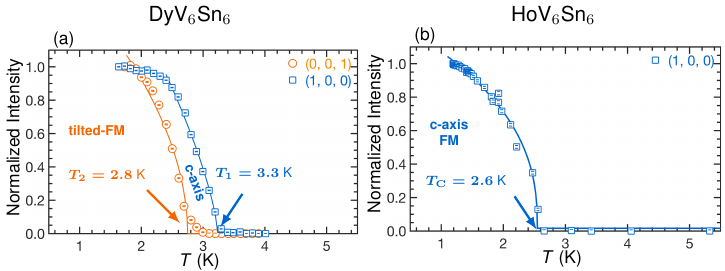}
\caption{(a) Normalized intensity $\left(I_{\mathrm{norm}} = \dfrac{I - I_{\mathrm{bckg}}}{I(1.6~\mathrm{K})}\right)$ as a function of temperature for the $(1,0,0)$ and $(0,0,1)$ Bragg peaks, illustrating the two magnetic phase transitions at $T_1 = 3.3(3)$~K and $T_2 = 2.8(5)$~K, respectively.(b) Normalized intensity as a function of temperature for the $(1,0,0)$ Bragg peak of \hvs\ illustrating single transition $T_c \sim 2.6$~K. Solid lines represent power-law fits.\label{OP}}
\end{figure*}

Additional neutron diffraction results presented in Fig.~\ref{OP} show the temperature dependence of the (1, 0, 0) and (0, 0, 1) Bragg peak intensities for \dvs\ and the (1, 0, 0) Bragg peak intensity for \hvs. These measurements were used to extract the temperature-dependent magnetic moment values presented in Fig.~2 of the main text.

\section{Mean-field calculations with anisotropic exchange \label{CAE}}
\begin{figure*}[]
\centering
\includegraphics[width=0.95\textwidth]{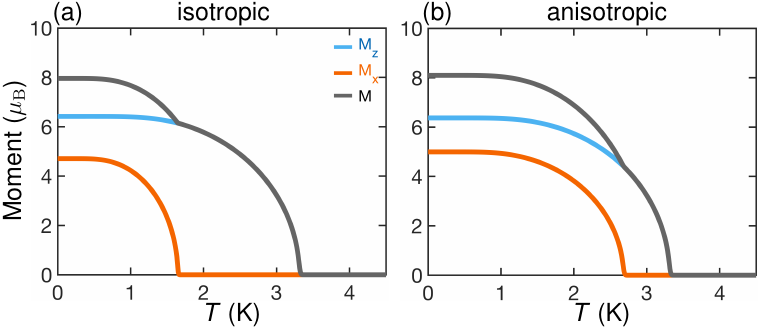}
\caption{(a) and (b) Magnetic moment as the function of temperatures from our mean-field calculations; (a) isotropic exchange as discussed in the main text, and (b) anisotropic exchange.\label{Iso_aniso}}
\end{figure*}

The mean-field calculations discussed in the main text were performed considering isotropic exchange interactions. Here, we extend the model by introducing anisotropic exchange interactions along the $x$ and $z$ directions, described by the Hamiltonian

\[
\mathcal{H}_{\mathrm{MF}}
=
\mathcal{H}_{\mathrm{CEF}}
-
z\left[
J_{\mathrm{ex}}^{xx}\langle J_x\rangle J_x
+
J_{\mathrm{ex}}^{z}\langle J_z\rangle J_z
\right],
\]

where $z$ is the coordination number. By considering stronger in-plane exchange interactions ($J_{\mathrm{ex}}^{xx} = 1.5 \times J_{\mathrm{ex}}^{z}$), we find that the transition temperature associated with the onset of the tilted FM phase shifts closer to the higher-temperature uniaxial FM transition, as shown in Fig.~\ref{Iso_aniso}, hence yields results in closer quantitative agreement with  the experimental transition temperatures. As discussed in the main text, long-range dipole-dipole interaction is potentially one of the reason for such an anisotropic exchange. 

\section{Spin-1 Blume-Capel model \label{BC spin-1}}
\begin{figure*}[]
\centering
\includegraphics[width=0.95\textwidth]{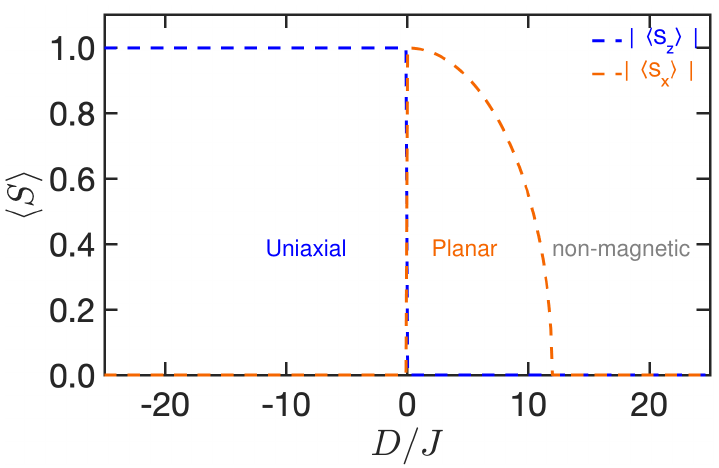}
\caption{Evolution of the spin expectation value $\langle S_\alpha \rangle$ as a function of the normalized single-ion anisotropy ratio $D/J$ in zero external field ($h = 0$), calculated using the spin-$1$ Blume--Capel model. The figure illustrates the sequence of transitions--from an initial uniaxial FM state to a planar, and subsequently to a nonmagnetic state.\label{BC_spin1}}
\end{figure*}

The Blume--Capel (BC) model, described by a spin-$1$ Hamiltonian with single-ion anisotropy (or crystalline electric field splitting), is defined as~\cite{BC_model}
\begin{equation}
    \mathcal{H} = -J \sum_{\langle i, j \rangle} S_i^z S_j^z - D \sum_i (S_i^z)^2 - h \sum_i S_i^z,
    \label{eq:blume_capel}
\end{equation}
where $S_i^z \in \{-1, 0, +1\}$ denotes the spin projection at site $i$, $J$ is the nearest-neighbor exchange coupling, $D$ is the single-ion anisotropy (CEF) parameter, and $h$ represents an external longitudinal magnetic field. Figure~\ref{BC_spin1} shows the evolution of the spin expectation value $\langle S^z \rangle$ as a function of the ratio $D/J$ at zero magnetic field ($h = 0$). Similar to the phase diagram shown for \hvs\ in Fig.~4 of the main text, the system undergoes successive phase transitions from a uniaxial to a planar configuration via a first-order transition, followed by a second-order transition into a nonmagnetic state. This provides further confirmation that \hvs\ can be effectively described as a spin-$1$ system within the Blume-Capel model.

\section{Large-exchange limit \label{LEL}}

\begin{figure}[]
\centering
\includegraphics[width=0.99\textwidth]{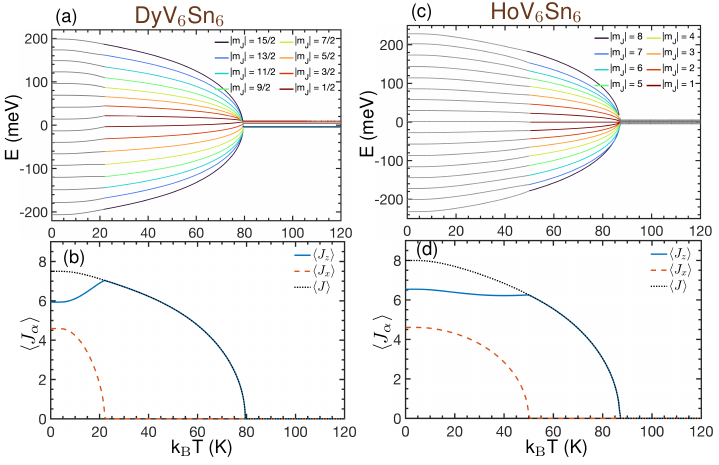}
\caption{\textbf{Large exchange limit calculations.}(a--b) for \dvs, and (c--d) for \hvs. (a) and (c) CEF levels calculated, for \dvs\ and \hvs, respectively, in the large-exchange limit with $J_0$ about 100 times larger than the exchange value used in our main text for small exchange limit. Solid lines denote states dominated by a single $\ket{m_J}$ eigenstates, while dotted lines indicate states with appreciable mixing between different $\ket{m_J}$ components. (b) and (d) Corresponding temperature dependence of the transverse, $\langle J_x\rangle$, and longitudinal, $\langle J_z\rangle$, components of the angular momentum.\label{Large_Jex}}
\end{figure}

Here, we test whether the system converges to the classical approximation in the large-exchange limit. Figure~\ref{Large_Jex} shows the evolution of the CEF levels for (a) \dvs\ and (c) \hvs, alongside the corresponding temperature dependence of the transverse, $\langle J_x\rangle$, and longitudinal, $\langle J_z\rangle$, angular momentum components in (b) and (d), respectively. These calculations are performed in the large-exchange limit, where the exchange coupling is set roughly $10^2$ times larger than the values used in the main text. In the regime $zJ_\mathrm{ex}\langle J \rangle^{2} \gg \Delta_{\mathrm{CEF}}$, the exchange interaction overcomes the crystal field, driving the system toward a fully polarized ground state with maximal angular-momentum projection, $\ket{J, m_J = J}$. Consequently, with increasing exchange field, the ground state rapidly evolves toward the $\ket{15/2}$ state in \dvs\ [Fig.~\ref{Large_Jex}(a)], and $\ket{8}$ state in \hvs\ [Fig.~\ref{Large_Jex}(c)]. Consistent with the classical MAE calculations presented in Fig.~1 of the main text, a canted phase with finite transverse magnetization, $m_x = g_J \langle J_x \rangle$, emerges at lower temperatures [Figs.~\ref{Large_Jex}(b) and (d)]. Thus, when the exchange interaction dominates over CEF effects, the quantum model naturally recovers the magnetic anisotropy predicted by the classical approximation.

\section{Irreps and eigenfunctions for Dy$^{3+}$ and Ho$^{3+}$ in hexagonal symmetry}
Given that both Ho$^3+$ and Dy$^3+$ have $D_{6h}$ hexagonal point-group symmetry, we adopt the irreducible-representation notation of Segal and Wallace
for the CEF eigenstates in hexagonal symmetry  discussed in \cite{Segal_1970}. 
Within this notation, the $J=15/2$ multiplet of Dy$^{3+}$ decomposes as
\[
3\Gamma_{7} + 2\Gamma_{8} + 3\Gamma_{9},
\]
whereas the $J=8$ multiplet of Ho$^{3+}$ decomposes as
\[
2\Gamma_{1}+\Gamma_{2}+\Gamma_{3}+\Gamma_{4}
+3\Gamma_{5}+3\Gamma_{6}.
\]
Here, $\Gamma_{7}$--$\Gamma_{9}$ are Kramers doublets for Dy$^{3+}$,
while $\Gamma_{1}$--$\Gamma_{4}$ are singlets and $\Gamma_{5}$ and
$\Gamma_{6}$ are non-Kramers doublets for Ho$^{3+}$. 

The symmetry-allowed basis states of the three irreducible representations for \dvs\ are
\begin{align}
\Gamma_{7}:&\;
\left\{
\left|\pm\frac{13}{2}\right\rangle,
\left|\pm\frac12\right\rangle,
\left|\mp\frac{11}{2}\right\rangle
\right\},
\\
\Gamma_{8}:&\;
\left\{
\left|\pm\frac72\right\rangle,
\left|\mp\frac52\right\rangle
\right\},
\\
\Gamma_{9}:&\;
\left\{
\left|\pm\frac92\right\rangle,
\left|\mp\frac32\right\rangle,
\left|\mp\frac{15}{2}\right\rangle
\right\},
\end{align}
where the opposite signs correspond to the two members of each Kramers doublet.
Following Segal and Wallace, the crystal-field eigenfunctions can be written in the general form,
\begin{align}
|\Gamma_{7}\rangle
&=
a\left|\pm\frac{13}{2}\right\rangle
+b\left|\pm\frac12\right\rangle
+c\left|\mp\frac{11}{2}\right\rangle,
\\
|\Gamma_{8}\rangle
&=
d\left|\pm\frac72\right\rangle
+e\left|\mp\frac52\right\rangle,
\\
|\Gamma_{9}\rangle
&=
f\left|\pm\frac92\right\rangle
+g\left|\mp\frac32\right\rangle
+h\left|\mp\frac{15}{2}\right\rangle,
\end{align}
where the coefficients are determined by diagonalizing the crystal-field
Hamiltonian.

Similarly, for the Ho$^{3+}$, the crystal-field eigenfunctions can be written as,
\[
\begin{aligned}
|\Gamma_{1}\rangle &= a|0\rangle+b\left(|+6\rangle+|-6\rangle\right), \quad
|\Gamma_{2}\rangle = \frac{1}{\sqrt{2}}\left(|+6\rangle-|-6\rangle\right), \\
|\Gamma_{3}\rangle &= \frac{1}{\sqrt{2}}\left(|+3\rangle+|-3\rangle\right), \quad
|\Gamma_{4}\rangle = \frac{1}{\sqrt{2}}\left(|-3\rangle-|+3\rangle\right), \\
|\Gamma_{5}\rangle &= c|\pm8\rangle+d|\pm2\rangle+e|\mp4\rangle, \quad
|\Gamma_{6}\rangle = f|\pm7\rangle+g|\pm1\rangle+h|\mp5\rangle .
\end{aligned}
\]

\end{appendices}

\bibliography{DyV6Sn6_References}

\end{document}